\documentclass[a4paper,twocolumn,twoside,11pt]{article}

\usepackage[T1]{fontenc}
\usepackage[utf8]{inputenc}
\usepackage[english]{babel}
\usepackage{indentfirst}
\usepackage{abstract}
\usepackage[hang,flushmargin]{footmisc}
\usepackage{titlesec}
\titleformat{\section}%
  {\centering\expandafter\MakeUppercase\expandafter}{\thesection.}{0.5em}{}
\titleformat{\subsection}%
  {\centering\slshape}{\thesubsection.}{0.5em}{}
\titleformat{\subsubsection}[runin]%
{\normalfont\normalsize\bfseries}{\thesubsubsection.}{1em}{\addperiod}
\newcommand{\addperiod}[1]{#1.}
\usepackage{caption}

\makeatletter
\newcommand{\nextauth}{%
 \@ifnextchar[{\nextauth@email}{\nextauth@nothing}%
}
\newcommand{\nextauth@nothing}[2]{%
  \phantom{x}#1$^{#2}$\hspace{-3pt}
}
\newcommand{\nextauth@email}[3][]{%
\if\relax\detokenize{#1}\relax
 \phantom{x}#2$^{#3}$
\else
 \phantom{x}#2$^{#3}$\thanks{\emailname: \texttt{#1}}\,\,%
\fi
}
\makeatother

\newcommand{\authors}[2][]{
   \author{\parbox{.9\linewidth}{\centering%
     \textbf{%
     #2}}}
   \markboth{\rm \hfill #1 \hfill}{}
   }

\newcounter{affil}
\newcommand{\nextaffil}[1]{%
  \stepcounter{affil}
  $^{\arabic{affil}}$\textit{#1}\\
}

\newcommand{\affiliations}[1]{
\date{#1}
}

\newcommand{\keywords}[1]{%
\phantom{x}\\
\noindent\textit{{\keywordsname}:} #1}
\newcommand{\doi}[1]{%
\phantom{x}\\
\noindent\textbf{DOI:} #1}

\def\emailname{E-mail}
\def\keywordsname{Keywords}

\newcommand{\titles}[2][]{
\title{\expandafter\MakeUppercase\expandafter{\bf{#2}}}
\markright{\expandafter\MakeUppercase\expandafter{\small\hfill\rm #1 \hfill}}
}

\newcommand{\wideabstract}[1]{
\twocolumn[
    \begin{onecolabstract}    
   \maketitle
   \noindent\normalsize
#1
\vspace{\baselineskip}
\end{onecolabstract}
]
\saythanks 
}

\def\aap{\rm{Astron.\ Astrophys.}}

\def\apj{\rm{Astrophys.\ J.}}

\def\araa{\rm{Ann.\ Rev.\ Astron.\ Astrophys.}}

\def\mnras{\rm{MNRAS}}

\usepackage{graphicx}
\usepackage{latexsym,amssymb,amsmath,color}
\usepackage{hyperref}
\usepackage{natbib}
\setcitestyle{aysep={,}}

\begin{document}
	
\newcommand{\gcc}{\mbox{g/cm$^{3}$}}
\newcommand{\XC}{X_\text{C}}
\newcommand{\rhoi}{\rho_\mathrm{ign}}
\newcommand{\rhom}{\rho_{9}}
\newcommand{\rhoim}{\rho_\mathrm{ign,9}}
\newcommand{\rhomin}{\rho_\mathrm{min}}
\newcommand{\tdiff}{t_\text{diff}}
\newcommand{\tdiffa}{t_\text{diff1}}
\newcommand{\tdiffb}{t_\text{diff2}}
\newcommand{\tnu}{t_{\nu}}
\newcommand{\ttrans}{t_\text{trans}}
\newcommand{\Ttrans}{T_\text{trans}}
\newcommand{\Tmax}{T_\mathrm{max}}
\newcommand{\Tign}{T_\mathrm{ign}}
\newcommand{\Etot}{E_\mathrm{tot}}
\newcommand{\Egam}{E_{\gamma}}
\newcommand{\Enu}{E_{\nu}}
\newcommand{\Qb}{Q_\mathrm{b}}
\newcommand{\msun}{\mbox{${M}_\odot$}}

\renewcommand{\abstractname}{\vspace*{-8ex}}
\renewcommand{\figurename}{{\small\textbf{Fig.}}}

\authors[KAMINKER et al.]{
  \nextauth[kam@astro.ioffe.ru]{A. D. Kaminker}{1},
  \nextauth{A. Y. Potekhin}{1},
  \nextauth{D. G. Yakovlev}{1}
}

\titles[neutron-star superbursts]{Neutrino emission of neutron-star superbursts}

\affiliations{
\nextaffil{Ioffe Institute, Politekhnicheskaya 26, St.~Petersburg, 194021 Russia}
}

\wideabstract{Superbursts of neutron stars are rare but powerful  events explained by
the explosive burning of carbon in the deep layers  of the outer envelope of the star.
In this paper we perform a  simulation of superbursts and propose a simple method for
describing  the neutrino stage of their cooling, as well as a method for describing  the
evolution of the burst energy on a scale of several months. We note  a universal
relation for the temperature distribution in the burnt  layer at its neutrino cooling
stage, as well as the unification of  bolometric light curves and neutrino heat loss
rates for  deep and powerful bursts. We point out the possibility of long-term 
retention of the burst energy in the star's envelope. The results can  be useful for
interpretation of superburst observations.
\doi{10.31857/S...}
} 


\section{Introduction}
\label{s:introduct}

Neutron stars (NSs) exhibit explosive activity of various types. In particular, the
bursts of accreting NSs are explained by explosive nuclear burning in their outer
envelopes. Comparison of observations with theoretical models allows one to obtain
useful information on the physical conditions in exploding layers, on accretion regimes,
masses and radii of the NSs, the equation of state of superdense matter in NS cores, and
much more \citep[see, e.g.,][]{Zand17,GallowayKeek21}.

Accreting NSs in binary systems demonstrate X-ray bursts caused by explosions of
accreted hydrogen and helium under the very surface of the star, as well as much rarer
but stronger superbursts (SBs) initiated by an explosion of carbon ($^{12}$C) in deeper
layers;  
see, e.g., a review by \citet{Zand17}.

In this paper we consider the SBs. They have been actively modeled for more than twenty
years with an account of accretion dynamics, chains of nuclear reactions, various
mechanisms of heat transport, neutrino cooling, etc.{} 
\citep[see, e.g.,][]{CummingMacbeth04,Cumming_06,KeekHeger11,Altamirano_12,Keek_12,Keek_15}.

SBs occur in deep layers of the outer envelope of an
NS \citep[e.g.,][]{HaenselPY07}. The envelope
extends from the NS surface to the neutron-drip
density of the matter ($\rho_\mathrm{drip} \approx 4.3 \times 10^{11}$
$\gcc$). The thickness of the envelope is several hundred meters, and
its mass is $\sim 10^{-5}$\,\msun. It consists mainly of
electrons and ions (partially or completely ionized atomic nuclei),
with the ions forming a crystal, liquid, or gas.

In particular, we are interested in neutrino cooling of the SBs. Previously, it
was included in calculations but was not studied systematically. In addition,
we will analyse the dynamics of the evolution of the SB energy over several
months and pay attention to the possibility of retaining the burst energy in
the star for a long time.

\section{The problem and numerical code}
\label{s:Problem}

It is well known \citep[e.g.,][]{KeekHeger11,Altamirano_12,Keek_12,Keek_15} that the  actual burst is
preceded by the stage of formation of carbon ($^{12}$C)  due to hydrogen-helium burning in the outermost
layers of an NS. The  matter containing carbon gradually accumulates and sinks into the  envelope under
the weight of newly accreted matter, and it reaches  explosive ignition conditions at the bottom of the
carbon layer with a  density of $\rhoi$ and temperature $\Tign$ \citep[e.g.][]{KeekHeger11}.  Following
the ignition, a rapid (lasting several minutes) nuclear  burning occurs in a wide carbon layer,
$\rhomin\leq\rho\leq\rhoi$,  where $\rhomin$ is a density at its outer boundary. The explosion power  is
determined by the parameter $\Qb$, equal to the produced  energy per one nucleon; $\Qb$ may vary
depending on the conditions of  the problem. According to calculations, a 
carbon-containing matter burns out during an explosion into  the iron group elements.

According to the theory, the maximum ignition density is limited by  $\rhoim\equiv \rhoi/(10^9~\gcc)
\lesssim 5$, because the $^{12}$C nuclei cannot survive at higher $\rho$ due to pycnonuclear reactions
and  beta-captures \citep[e.g.,][]{ShapiroTeukolsky}.

After the explosion, thermal relaxation of the heated layer occurs, observable in the light
curves for several hours or days. However, the internal thermal relaxation can last more than a year 
\citep[e.g.,][]{KeekHeger11}. It is determined by the diffusion of the  released heat to the surface and
inside the NS, and by neutrino cooling  of the heated matter.

To simulate an SB, we used a numerical code developed for studying the  cooling of NSs on various
spatial and temporal scales, which includes modern microphysics of the  NS matter
\citep{PotekhinChabrier18}. A  feature of this code, which manifests itself when modeling the 
propagation of heat in the outer envelopes of NSs for relatively short  times, is an accurate account of
the removal of electron  degeneracy with decreasing density or increasing temperature.  The thermal
evolution of the entire star is calculated, although the  temperature of the core 
is almost unchanged
during observations. We  do not pretend to perform a complete self-consistent SB modeling, but  make
several simplifications, whose validity is discussed below.

We do not study the evolution of accreted matter in the NS envelope  before the burst, such as the
dynamics of ignition, formation and  propagation of a shock wave, and related phenomena 
\citep[e.g.,][]{KeekHeger11}. We assume that the burst occurs in a  layer $\rhomin\leq\rho\leq\rhoi$,
where $\Qb$ does not depend on density  $\rho$. The quantities $\rhomin$, $\rhoi$, $\Qb$, and $\Tign$
are  treated as free parameters. The SB energy is mainly released  near the ignition density
$\rhoi$; hence the results are  insensitive to $\rhomin$, if $\rhomin\ll\rhoi$ (for certainty, we
fixed  $\rhomin = 10^7$ \gcc). The initial temperature profile corresponded to  the thermal
quasi-equilibrium in a cooling NS at $\Tign \sim (1-5)  \times 10^8$~K. At any density $\rho$ in the
chosen layer, the burst  energy was released uniformly over time for 100 seconds (in the local 
reference frame), after which the energy release stopped. We varied  $\rhoim$ from $0.1$ to $5$. 

The values introduced above are determined in the local reference frame of the NS outer envelope.
The values in the system of a remote observer will be hereafter marked by a tilde. We emphasize that
everywhere (in the figures and in the text), except for Section \ref{s:approx},  $t$ means
coordinate Schwarzschild time. In Section \ref{s:approx}, however, the time in the local reference frame
of the NS envelope is introduced; this is taken into account when comparing with the results of 
other sections.

\section{Pure neutrino cooling}
\label{s:approx}

It is convenient to describe the initial post-explosion period of the cooling of the heated layer in a
simple approximation of pure neutrino cooling. It is sufficient to use the well-known approximation
of instantaneous explosion \citep[e.g.,][]{Altamirano_12} assuming the temperature in the
burning layer at any density $\rho$ adiabatically reaches its maximum value $\Tmax(\rho)$. 
Then the
neutrino emission cools any matter element independently according to the
equation
\begin{equation}
C \frac{\partial T}{\partial t}=-Q_\nu,
\label{e:nucool}	 
\end{equation}  
where $C$ is the heat capacity (at constant pressure $P$) per unit  volume, and $Q_\nu$
is the neutrino energy production  rate in this volume. The main neutrino processes in  SBs are the
neutrino emission  due to Langmuir plasmon decays and the annihilation of  electron-positron pairs into
neutrino pairs. The solution of  Eq.~(\ref{e:nucool}),
\begin{equation}
t=\int_{T}^{T_\mathrm{max}} \frac{C(T')}{Q_\nu (T')}\; d T',
\label{e:nucool1}
\end{equation}
depends (parametrically) only on the values of $\rho$ and $\Qb$ in the given matter element; 
therefore, 
$T=T(t,\rho,\Qb)$. This solution is not sensitive to an NS model (mass,  radius, equation of state) and
radically simplifies the study of the  neutrino stage of an SB cooling. However, when heat diffusion
becomes  important \citep[see, e.g.,][]{Yakovlev_21}, the noted universality is  violated.

Let us add that, when modeling SBs, the column depth of matter $y$ (g/cm$^2$) is often used instead of
the mass density $\rho$, and the column depth of the burst energy $E_\mathrm{i}$ (erg/cm$^2$) is often
used  instead of $\Qb$. But the  solution (\ref{e:nucool1}) is insensitive to the NS model just when 
using the quantities $\rho$ and $\Qb$. One such solution is easily  applicable to any model. In this
sense, the temperature distributions  at the neutrino stage of a SB are universal for different NS
models.

This property is useful for testing calculations carried out with a computer code. In addition, the
solution (\ref{e:nucool1}) can in principle be used in the code itself to speed up calculations at the
neutrino stage of an SB.

\begin{figure}[th!]
	\centering
	\includegraphics[width=\columnwidth]{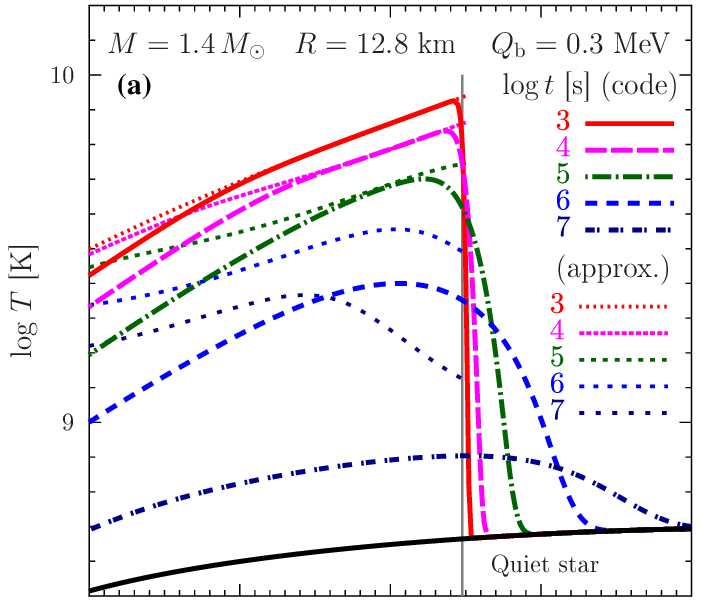}
	\includegraphics[width=\columnwidth]{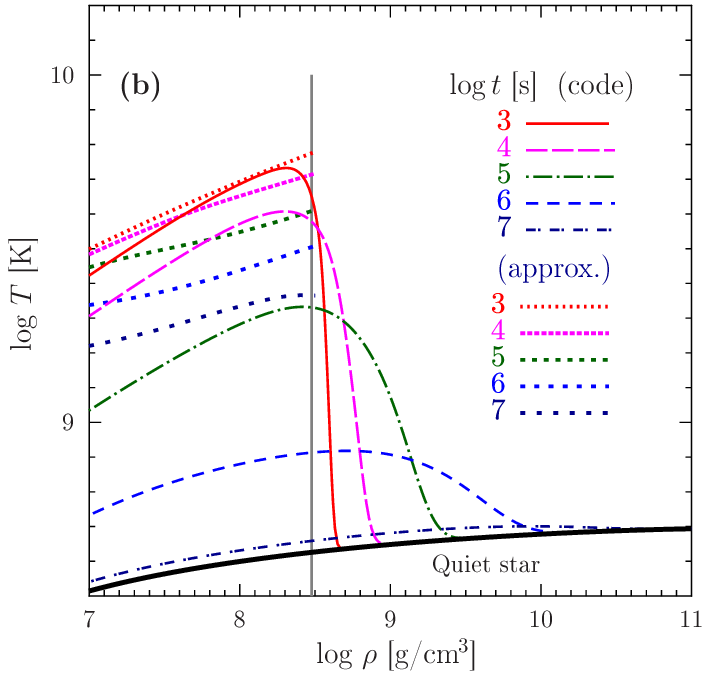}
        \caption{The internal temperature $T$ 
        versus mass density $\rho$ in the burst
        area at different times $t$ for $\Qb$=0.3 MeV and two ignition depths:  
        (a) $\rhoim=3$ (thick curves) and (b) $\rhoim=0.3$ (thin curves). 
        The dotted
        lines show the neutrino cooling approximation; they are truncated  at
        $\rho > \rhoi$ (at $\rho_9 < 0.3$ the dotted curves in the
        cases (a) and (b) coincide).
	}
	\label{fig1}
\end{figure}

\begin{figure}[t]
	\centering
	\includegraphics[width=\columnwidth]{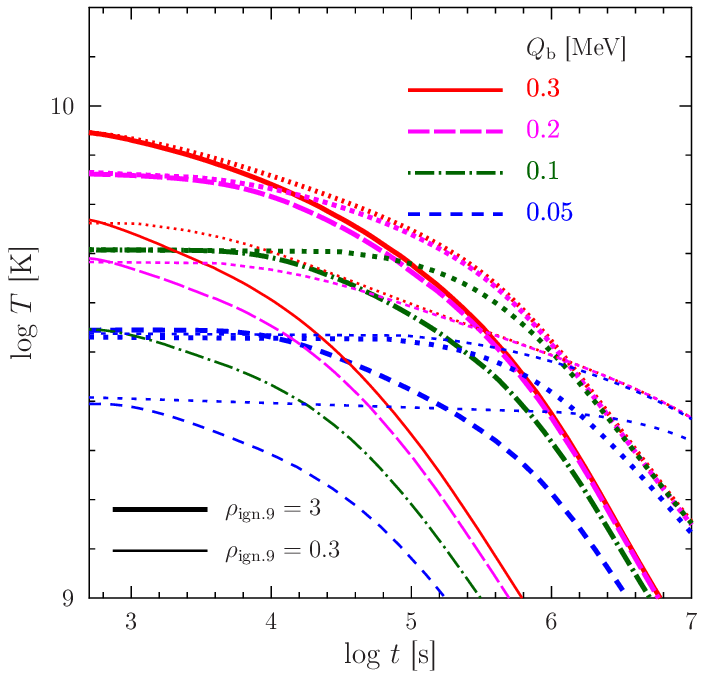} 
	\caption{The dependence $T(t)$ for the two ignition models in 
	Fig.~\ref{fig1} in the layers with $\rho=0.8 \, \rhoi$ at 
	$\Qb=$0.3, 0.2, 0.1, and 0.05 MeV. Solid, dashed, and dot-dashed 
	lines correspond to the numerical simulations; dotted lines show the neutrino 
	cooling approximation. Here and hereafter,  the end of 
	heating in the layer $(\rhomin,\rhoi)$ is taken as the starting 
	point $t=0$.
	}
	\label{fig2}
\end{figure}

\begin{figure}[h!]
	\centering
	\includegraphics[width=\columnwidth]{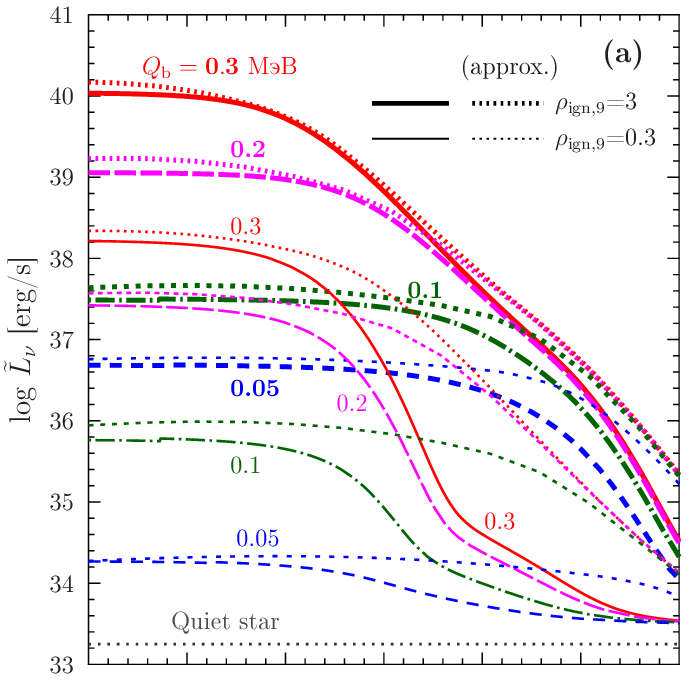}   
	\includegraphics[width=\columnwidth]{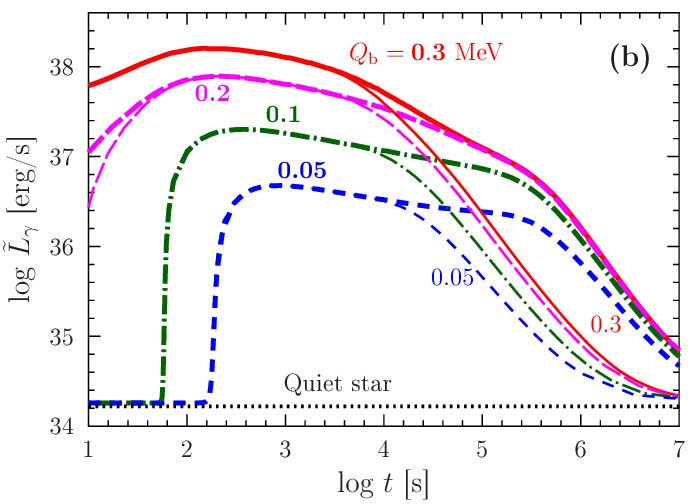}  
	\caption{Neutrino $\widetilde{L}_\nu(t)$ (a) and electromagnetic 
	$\widetilde{L}_\gamma(t)$ (b) light curves, calculated for the same 
	SB models as in Fig.~\ref{fig2}. The luminosities are redshifted
	(for a distant observer).
}
	\label{fig3}
\end{figure}

\section{Results}
\label{s:results}

We present the results of SB modeling with the numerical code and  compare them, where possible, with
the approximation of pure neutrino  cooling. In the figures below (except for Fig.~\ref{f:U}) we have
used  a model of an NS with mass of $1.4\,\msun${\bf.} The inner crust 
and the core are described by the
equation of state BSk24 \citep{Pearson_18},  and the outer crust consists of iron $^{56}$Fe (the radius
of such a star  is 12.8 km); we have also considered other models and obtained similar  results. 

Fig.~\ref{fig1} shows the profiles $T(\rho)$ in and near the exploded layer  for two SBs at different moments
of time $t \leq 10^7$~s in the  reference frame of the outer envelope. Fig.~\ref{fig1}(a) corresponds 
to $\rhoim =3$, and Fig.~\ref{fig1}(b) to $\rhoim =0.3$ at the same $\Qb=0.3$
MeV. The curves are marked with the values  of log\,($t$~[s]). Upper curves ($t=1000$~s, almost
immediately after  the SB) are close to the thermodynamic curve $\Tmax(\rho)$, which is  the same at a
fixed $\Qb$. Note the sharp breaks in $\Tmax(\rho)$ at  $\rho\approx\rhoi$. The spreading of heat from
the explosion area  gradually turns the break into a blurred maximum, with the heat  diffusing into the
NS at larger depths and outwards at lower $\rho$.

The dotted lines in Figs.~\ref{fig1}(a) and \ref{fig1}(b) show
the  temperature profiles  for pure
neutrino cooling. In case (a) of deep  ignition ($\rhoim=3$), this approximation works well in the
layer where the main burst energy has been produced ($\rho_9 \gtrsim 1$) during $t  \lesssim (1-3) \times
10^5$~s. With the pure neutrino cooling, the  temperature profiles $T(\rho,t)$ are universal and can be
easily  obtained without using the cooling code. In case (b) of a shallower ignition ($\rhoim=0.3$), the neutrino
cooling stage turns out to be much shorter ($t \lesssim 10^3$~s) and practically insignificant (see
below).  Thus, the approximate solutions coincide with the exact ones in 
the cases in which the cooling of
matter is determined by neutrino radiation, i.e.{} at sufficiently short times 
of deep and powerful bursts.

When the temperature  $T(\rho)$ drops, the neutrino cooling approximation is violated. So, in a layer
with an arbitrary density $\rho$, the neutrino cooling becomes insignificant at sufficiently low $\Qb$, at
which $T(\rho) \lesssim T_\nu$. Here $T_\nu \sim (3-4) \times 10^9$~K 
is a characteristic temperature that is weakly dependent
on $\rho$.  This circumstance has been repeatedly noted in the literature
\citep[e.g.,][]{CummingMacbeth04,Cumming_06}.

As an example, in Fig.\ \ref{fig2} we show the evolution of temperature  $T(t)$ in four SBs with
different heat releases ($\Qb=0.05-0.3$~MeV) and  with ignition either at $\rhoim=3$  (thick lines), or
at $\rhoim=0.3$  (thin lines). At each value of $\rhoi$, the profiles $T(t)$ are shown  for the layer
with density $\rho=0.8\,\rhoi$. 

It is remarkable that the cooling curves $T(t)$ for the two most powerful SBs ($\rhoim=3$, $\Qb\gtrsim
0.2$) at $t\gtrsim 10^4-10^5$~s merge into the same ``universal'' curve. This effect is well known in
the theory of neutrino cooling of 
NSs \citep[e.g.,][]{YakovlevPethick04} as the   ``memory loss'' of initial conditions because of a strong dependence of the neutrino cooling rate on
temperature: a noticeable excess of $T$ over $T_\nu$ causes rapid neutrino cooling down to $T\sim
T_\nu$, which ``unifies'' the cooling. If we increase $\Qb$ to still higher values, then the additional
burst energy will be carried away by neutrinos, and the dependence $T(t)$ at $t\gtrsim 10^5$~s will not
change.

However, in other cases, considered in Fig.~\ref{fig2}, the neutrino  cooling (at $\log t\gtrsim 4$) turns
out to be so weak that it does not actually affect $T(t)$.

Figs.~\ref{fig3}(a) and~\ref{fig3}(b) present bolometric neutrino and  electromagnetic light curves,
$\widetilde{L}_\nu(t)$ and  $\widetilde{L}_\gamma(t)$, for the same SB models as in  Fig.~\ref{fig2}. A
comparison of $\widetilde{L}_\nu(t)$ and  $\widetilde{L}_\gamma(t)$ shows that  for the shallower 
depth ($\rhoim=0.3$)  the photon luminosity $\widetilde{L}_\gamma(t)$  dominates at all the chosen
values of $\Qb$ (except for the earliest moments after  the nuclear burning). In the deeper burst
($\rhoim=3$), especially with high $\Qb$   ($0.3$ and 0.2 MeV), on the contrary,
the neutrino  heat   
outflow  
becomes decisive. Merging of the curves $\widetilde{L}_\nu(t)$  {at $\Qb = 0.3$ and
$\Qb =0.2$} confirms the conclusion on the  unification of cooling of deep and powerful SBs at
$t\gtrsim 3 \times  10^4$~s as a result of powerful neutrino emission. 

Dotted lines in Fig.\ \ref{fig3}(a) show the dependences  $\widetilde{L}_\nu(t)$, calculated in the
approximation of pure  neutrino cooling. They are obtained by integrating neutrino emissivities
$Q_\nu(\rho,t)$ over the volume of the exploded layer in the NS  envelope. A comparison with the results of
accurate calculations  confirms the conclusions on the duration and efficiency of the  neutrino
cooling stage 
drawn from
the analysis of Figs.~\ref{fig1}  and \ref{fig2}.

The  electromagnetic light curves shown in Fig.~\ref{fig3}(b) are formed after the heat wave
from the burning layer has  reached the surface  \cite[e.g.,][]{CummingMacbeth04,KeekHeger11,Keek_12}.
The initial segments of these curves are determined by the outer parts of this layer ($\rho \sim  \rho_\mathrm{min}$),
while 
later parts of the light 
curves are determined by deeper layers where most of the
energy is released ($\rho \sim \rhoi$). However, at the
early stage, the  dependences $T(\rho)$ at the same $\Qb$ are 
rather
close to each other  (Fig.\  \ref{fig1}). 
Therefore, the early segments ($t \sim 10^2-10^4$~s)  of the light curves
$\widetilde{L}_\gamma(t)$ for the SBs with such $\Qb$  
weakly differ from each other (Fig.~\ref{fig3}(b)).
The bursts with $\rhoim=0.3$ are shorter than the deeper bursts with $\rhoim=3$. The similarity of the
curves $\widetilde{L}_\nu(t)$, as well as of the curves $\widetilde{L}_\gamma(t)$, for the deeper SBs at
$\Qb=$0.3 and 0.2 MeV at  late stages ($t \gtrsim 10^4$~s) is explained by the same 
unifying influence
of powerful 
neutrino
heat 
outflow
at the beginning of SBs.

\begin{figure}[t]
	\centering
	\includegraphics[width=\columnwidth]{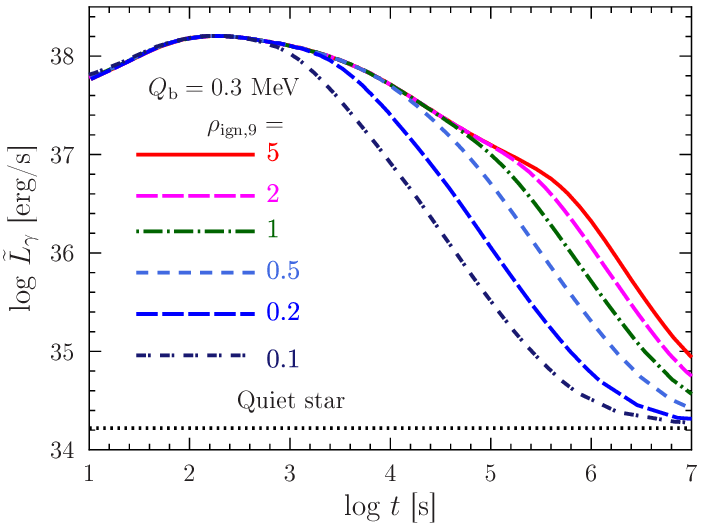}     
	\caption{
		Photon light curves $\widetilde{L}_\gamma(t)$ for superbursts,
		calculated for $\Qb=0.3$ MeV, at different ignition depths
		$\rhoim$ from 0.1 to 5. 
}	 
	\label{fig4}
\end{figure}

Fig.~\ref{fig4} presents the light curves $\widetilde{L}_\gamma(t)$,  computed for the bursts with
$\Qb=$0.3 MeV at six ignition depths  $\rhoim$ from 0.1 to 5. At $t\lesssim1000$~s, all the curves
almost coincide; later they differ but have common properties. We can see that with the growth
of $\rhoi$, the duration of the afterglow of the SB increases, which is associated 
with prolonging heat diffusion  from deeper layers to the  NS surface \citep[e.g.][]{Yakovlev_21}. 

Note that our calculations allow us to study the evolution of the  thermal energy of an SB. The
total energy of the SB, $\Etot$, is released  already a few minutes after the SB start in a layer of
burnt  carbon. Having known from the simulations, the photon $L_\gamma (t)$ and  neutrino  $L_\nu(t)$
luminosities of the NS, as well as luminosities  $L_{\gamma 0}$ and $L_{\nu 0}$ in the quiescent NS
(before the SB), we  find the luminosities $L_{\gamma}^*=L_{\gamma}-L_{\gamma 0}$ and 
$L_{\nu}^*=L_{\nu}-L_{\nu 0}$, which are determined by the SB  itself. Integrating them over time from
the start of the heating to the  current moment $t$, we obtain the photon and neutrino energies of the 
SB, $\Egam^*(t)$ and $\Enu^*(t)$, that have been taken away from 
the  star by any current time moment $t$.
The quantity $E_T^*(t)=\Etot-  \Egam^*(t)-E_\nu^*(t)$ can be called residual thermal energy of the SB in 
the NS. Here, the energies and luminosities are given in the local reference frame, since all the main
processes of thermal evolution occur in the outer crust, which is so thin that space-time is almost flat
in it, although with different time and spatial scales than in the system of a remote observer.

\begin{figure*}[t]
	\centering
	\includegraphics[width=.79\textwidth]{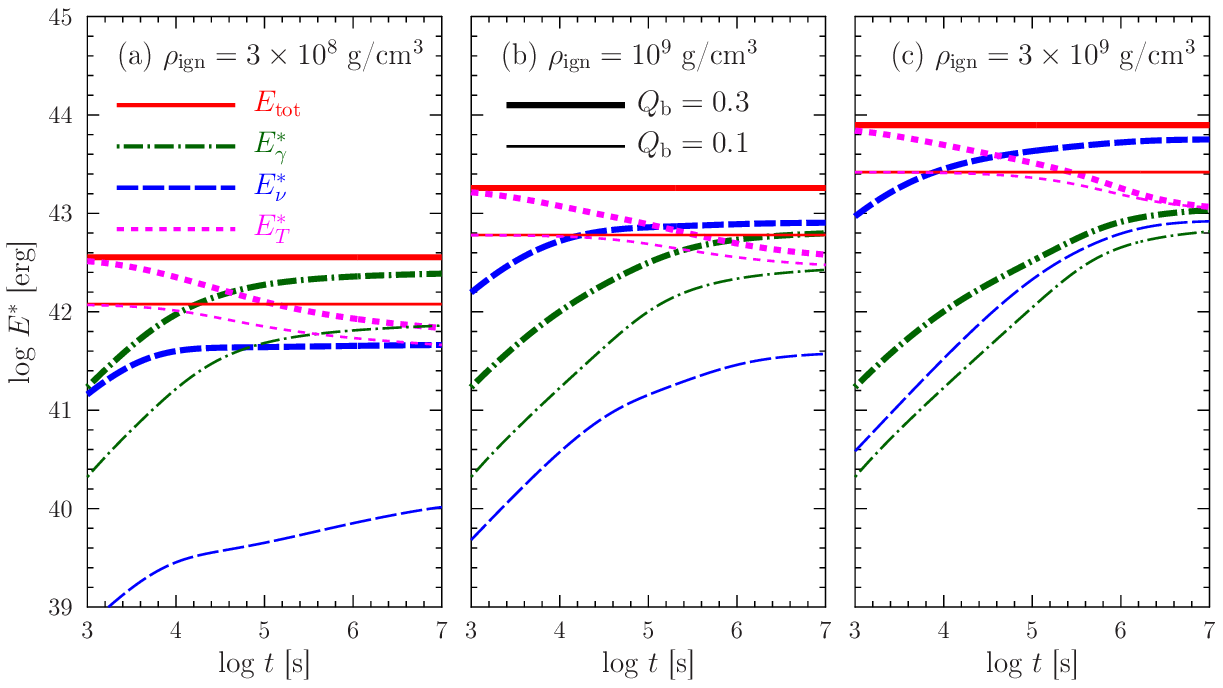} 
	\caption{The evolution of the thermal energy of SBs in time, 
	calculated at $\rhoim$=0.3 (a), 1 (b), and 3 (c) in the cases of 
	$\Qb$ = 0.3 MeV (thick lines) and 0.1 MeV (thin lines): 
	$E_\gamma^*$ is the energy of the SB that is radiated through the 
	star's surface, $E_\nu^*$ is the radiated neutrino energy of the 
	SB, $E_T^*$ is the residual energy of the SB in the NS. The solid 
	horizontal lines show the full energy of the burst $E_\mathrm{tot}$.
}
\label{f:triptix}
\end{figure*}

Fig.~\ref{f:triptix} gives examples of time-dependences $\Egam^*(t)$,  $\Enu^*(t)$, and $\Etot^*(t)$
for six SBs at three values of $\rhoim=  0.3$ (a), 1 (b), and 3 (c). Two bursts are considered for each
ignition  density, with sufficiently low ($\Qb=$0.1 MeV) and high ($\Qb$= 0.3  MeV) 
fuel efficiencies. The solid horizontal lines show full energies of the SBs. The active phase of the SB energy
removal continues longer for higher $\rhoi$.

Fig.~\ref{f:triptix}(a) shows the dependences for  shallower bursts.  We see that, at a weak SB
($\Qb=$0.1 MeV), the  neutrino cooling is negligible. The main relaxation mechanism of
such SB is  the emission of photons from the surface. However, even after $10^7$~s, when  the afterglow from
the surface is hardly observable, there remains over  $10^{41}$ ergs of the SB energy in the NS. A more
powerful burst  ($\Qb=0.3$ MeV) at this depth is accompanied by considerably stronger  neutrino cooling,
which, nevertheless, is comparable in intensity with the photon cooling during $\sim1000$~s, 
and weakens afterwards. Meanwhile  the photon cooling remains powerful, and in $10^7$~s it scoops out a 
significant fraction of $\Etot$.  

Fig.~\ref{f:triptix}(b) corresponds to two SBs at a moderate depth  $\rhoim$=1. The neutrino 
cooling
of the weaker burst ($\Qb= 0.1$ MeV) is inefficient, so that the most of the energy is carried out by
radiation through the surface. By the moment $t = 10^7$~s, there still remains $\sim  10^{42}$ ergs of
the SB energy in the NS. At a stronger burst ($\Qb=0.3$~MeV),  on the contrary, the excess energy is
carried away mainly by neutrinos, but the residual energy remains very large.

Fig.~\ref{f:triptix}(c) demonstrates two SBs 
with the ignition
depth of  $^{12}$C, which is close to the
limiting one. The weaker SB  differs from the weaker SBs in Figs.~\ref{f:triptix}a and 
\ref{f:triptix}b: now the neutrino heat loss dominates over the heat loss  through the surface, although
the difference between the heat losses through these two channels is small. Meanwhile, the residual 
energy by the moment $t=10^7$~s is only a few times smaller than the total  energy $\Etot$. Our
estimates show that in this case the main part of this energy 
flows into
the star.

Finally, the stronger burst ($\Qb=0.3$ MeV) in Fig.~\ref{f:triptix} is  accompanied by an unusually
powerful neutrino heat 
outflow.
This burst  belongs to a special class of carbon superbursts, which are
almost  entirely controlled by neutrino processes. According to the  calculations, the rate of heat
transport inside the star significantly  exceeds the rate of energy outflow through the surface. The
possibility  of implementing such bursts in the NSs is not yet clear.

As an example, in Fig.~\ref{f:U} we have depicted the thermal evolution of the
SB of the transiently accreting NS in the low-mass X-ray binary system 4U 1608--52. 
The burst was observed on May 5, 2005, by the instrument ``All-Sky Monitor'' 
onboard the \textit{Rossi X-ray Timing Explorer} space observatory. 
The analysis of the observations \citep[e.g.,][]{Keek_08, Zand17} shows that
it was apparently one of the most powerful and deepest superbursts. For 
an NS with $M=1.4\,\msun$ and $R=10$ km,
fitting of the observed X-ray light curve by theoretical
models
gave (in our notation)
 $\rhoim \approx 1.3$,
$\Qb \approx 0.17$ MeV.

Fig.~\ref{f:U} presents results of our calculations for this burst (with the indicated NS parameters) in
the same form as in Fig.~\ref{f:triptix}.  The neutrino radiation effectively cools the SB for a relatively
short time. For several hours, the  neutrino cooling rate is (coincidentally) close to the rate of
photon emission through the NS surface. Later, the photon cooling from the NS surface dominates, 
but by
 $t = 10^7$~s (about 4 months) it becomes weak
too. During this time, the neutrinos 
carry away about a quarter  of the total energy of the burst and the photons radiate away about 60\% of it.
The residual thermal energy is $\gtrsim 10^{42}$~erg.

These results, as well as the results presented in Fig.~\ref{f:triptix}, indicate that even a few months
after an SB, a  significant part of the explosion energy may remain in the star.  Calculations show that
this energy (especially for deep SBs) largely  flows into the deep layers of a NS. The question of the
further history  of this energy is not trivial. For example, in the presence of  regularly recurring
bursts, their heat gradually warms up the  entire star. It is possible to choose the frequency or
intensity of the  bursts so that, in general, the star remains in a quasi-stationary  state
\citep[e.g.,][]{Colpi_01,KeekHeger11}. The evolution  of residual heat can be determined by small and
rather complex  temperature gradients in the NS envelope, which can appear there for various  reasons.
This issue requires further study.

\begin{figure}[t]
	\centering
	\includegraphics[width=\columnwidth]{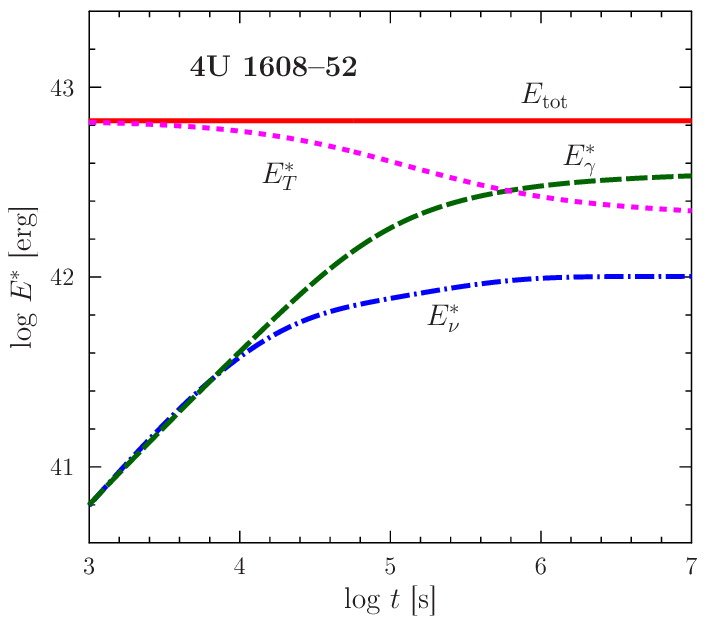} 
	\caption{
The evolution of thermal energy of the powerful NS burst in the X-ray source 4U 1608--52,
calculated for SB parameters taken
from interpretation of its observations.
} 
	\label{f:U}
\end{figure}


In addition, we have performed a series of simulations of  NS superbursts with different
equations of state of superdense matter,  with varying the mass (and radius) of the star, as well as other
parameters  of the problem, including $\rhomin$ and $\Tign$. The use of different  equations of state
and NS parameters did not reveal qualitatively new  effects. Reducing $\rhomin$ below the adopted value
$10^7$ \gcc\  \citep[which is quite acceptable; see, e.g.,][]{KeekHeger11,Keek_12,Altamirano_12} cannot
noticeably change the global SB evolution. It can affect very early segments of photon light curves,
which we do not model anyway. Variations of the ignition temperature $\Tign$ also have a weak effect on
the SB model as long as this temperature is noticeably lower than the temperature of the heated matter
after the explosion. The ignition temperature can be determined self-consistently for each individual
burst. However, a self-consistent calculation is complicated and requires not always reliably known
microphysics (for example, the rates of many reactions that accompany the carbon burning). The
conclusion that the results of our simplified modeling are not sensitive to the exact values of $\Tign$
confirms the adequacy of our approach.

Whenever possible, we tried to compare our results with the results of other authors. In particular, we
obtained a good agreement with fig.~2 in the paper by \citet{Cumming_06}, where the
neutrino and photon cooling rates after a burst were compared, and with fig.~5 in the paper by
\citet{KeekHeger11}, where the dependences  $L_\gamma(t)$ and $L_\nu(t)$ were presented for one of SBs.

\section{Conclusion}
\label{sect:concl}

We have considered the main features of rare and very energetic events, the superbursts, occurring in
deep layers of the outer envelopes of accreting NSs; they are caused by the explosive combustion of
accumulated carbon ($^{12}$C) in these 
layers. For their modeling, we used a numerical code
\citep{PotekhinChabrier18} based on modern NS microphysics (Section~\ref{s:Problem}),
introduced the approximation of pure neutrino cooling (Section~\ref{s:approx}), and analysed the dynamics
of heat propagation during the post-burst relaxation (Section~\ref{s:results}).

A simple method for studying the neutrino cooling stage of the outer crust is  proposed, which allowed
us to formulate a universal  relation for the temperature distribution $T(t, \rho, \Qb)$ in 
the exploded 
layer.

It is noted that the dynamics of the SBs is determined by two main parameters: the density $\rhoi$ of
explosive ignition of carbon and the  
fuel energy $\Qb$ deposited
per nucleon. At rather shallow SBs ($\rhoim \lesssim
0.3$) the neutrino cooling turns out to be insignificant for any considered $\Qb$; one can call such SBs
``neutrinoless''. With increasing $\rhoi$  at sufficiently 
high $\Qb$, the neutrino cooling is beginning to
be important. The duration of neutrino cooling stage is increasing (from $\sim 10^3$~s  at
$\rhoim \sim 0.3$ to $\gtrsim 10^5$~s at $\rhoim \sim 3$). With growing $\Qb$ to  $\sim 0.3$
MeV, the SB-generated neutrino luminosity $L_\nu(t)$ starts to exceed the photon luminosity 
$L_\gamma(t)$ from the NS surface, taking control of the cooling. At especially deep ignitions $\rhoim
\gtrsim 3$ and high  $\Qb \sim 0.3$ MeV, the main fraction of the SB energy is carried away
by neutrinos, and the light curve $L_\gamma(t)$ is determined by only a small fraction of the SB energy.
Moreover, with increasing $\rhoi$, more and more energy is carried by thermal conduction inside the
NS. In certain cases, a fairly large burst energy can remain in the star even a few months after the SB.

The employed cooling code facilitates simulations of the evolution of burst energy over time
(Fig.~\ref{f:triptix}). A comparison with observations can help one to study the dynamics of SB
energy for specific events  (Fig.~\ref{f:U}).

Once again, we emphasize that SB modeling has been performed at an advanced level for more than 20 years
\citep[see, e.g.,][and references
therein]{CummingMacbeth04,Cumming_06,KeekHeger11,Altamirano_12,Keek_12,Keek_15,Zand17,GallowayKeek21},
Many of the results mentioned above (for example, the possibility of removing 
a large fraction the SB energy
by neutrinos) were obtained earlier \citep[e.g.,][]{Cumming_06,KeekHeger11}. Our consideration allowed
us to determine a number of general 
features
of the SBs. 
These include: a simplified description of the
neutrino cooling stage of the outer crust; 
the possibility of studying the evolution of SB thermal
energy on scales of several months 
and the conclusion on possibility of retaining SB energy
inside the star on such time scales.

The maximal ignition depths $\rhoim \sim 5$ 
considered here are perhaps limiting
for the explosive burning of $^{12}$C \citep[see, e.g.,][]{PotekhinChabrier12}.
However, one cannot exclude that bursts due to burning of matter with
different composition 
are possible in still deeper layers. For instance, \citet{Page_22} suggested that
the powerful bursting activity of the X-ray source
MAXI J0556--332 is related to a hyperburst in an NS at
$\rhoim \sim 100$ (near the boundary of the outer envelope and inner crust) 
as the effect of explosive burning of a number of isotopes with large neutron excesses.
According to the theory, such hyperbursts, if they occur, are very rare.
If they do occur, they represent a continuation of the family
of ordinary carbon superbursts, but in an extremely unusual mode.

Our results can also be useful
for studies of conventional thermonuclear bursts in  the surface layers of NSs.
First of all, these are the so-called intermediate duration bursts, 
which are shorter than the SBs. They occur in a helium layer
on cold NSs \citep[e.g.,][]{Cumming_06,Zand17}.
Similar energy transfer problems occur with SBs or ordinary bursts of magnetars
in their inner or outer crusts under the influence of superstrong magnetic fields
\citep[e.g.,][]{KaspiBeloborodov17}.

The work was supported by the 
Russian Science Foundation (grant 19-12-00133-P).

\renewcommand{\refname}{References}


\end{document}